\documentclass[twocolumn,10pt]{article}
\usepackage{sidecap,graphicx} 
\title{Spectra of particles produced in high-mass diffraction dissociation 
in the Model of Quark-Gluon Strings}
\author{A.B. Kaidalov$^1$ and M.G. Poghosyan$^2$\\
$^1$Institute of Theoretical and Experimental Physics, 117259 Moscow, Russia\\
$^2$Universit\'a  di Torino/INFN, 10125 Torino, Italy
}                     
\begin{document}
\maketitle
\begin{abstract}
\vspace{-0.3 cm}
We calculate spectra of secondary particles produced in process 
of single diffraction dissociation in the Model of Quark-Gluon 
Strings. For description of diffractive dissociation process we 
use the reggeon model where all legs of triple-reggeon diagrams 
are eikonalized. Numerical calculation shows that the model gives 
a good description of data on charged particles pseudorapidity 
distribution in $p\bar{p}$ single diffractive dissociation.
\end{abstract}
\vspace{-0.5 cm}
\section{Introduction}
\vspace{-0.3 cm}
\label{intro}
The role of hardonic soft interactions is hard to overestimate, 
since they are responsible for multi-particle production in 
high-energy hadronic collisions. Low-momentum transfer characterizes 
these processes and the strong interaction constant is not a small 
parameter for them. As a consequence pQCD cannot be used for their 
study. The models of Quark-Gluon Strings  (QGSM) \cite{QGSM} and Dual-Parton 
Model (DPM) \cite{DPM}, are based on nonperturbative notions, combining $1/N$ 
expansion in QCD with Regge theory and using parton structure of hadrons. In 
both approaches the matrix element of hadron-hadron scattering amplitude 
consists from planar diagrams (related to flavor exchange), which are 
associated with secondary-Reggeon exchange diagrams in Reggeon Field Theory 
(RFT) and cylinder-type diagrams (related to color exchange), which are 
associated with Pomeron exchange in RFT. 

$1/N$ expansion is a dynamical expansion and the speed of convergence depends 
on the kinematical region of studying process. At very high energies many terms 
of the expansion (multi-pomeron exchanges) should be taken into account. 

In ref. \cite{SDXS} a model has been developed, where all possible eikonal-type 
corrections are taken into account in triple-reggeon diagrams. It was applied for 
discription of soft single-diffractive dissociation process in hadronic interactions. 
It was shown that it is possible to describe data on diffractive $pp$ and $p\bar{p}$  
differential cross-sections in a wide energy range (from $P_{lab}$ = 65 GeV/c to 
$\sqrt{s}$ = 1800 GeV) accessible by different accelerators of CERN and Fermilab. 
In this article we incorporate QGSM into the model and give description of data on 
spectra of secondary particles produced in single diffraction dissociation. 
\vspace{-0.5 cm}
\section{Single-diffraction dissociation cross-section}
\begin{SCfigure}[][h]
\centering
\resizebox{0.3\textwidth}{!}{%
\includegraphics{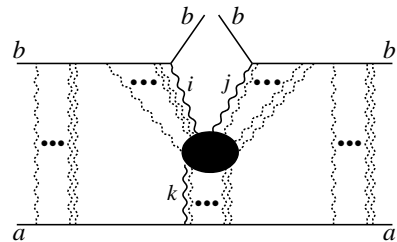}
}
\caption{
Eikonalized triple-reggeon diagram that is proposed in Ref. \cite{SDXS}
to describe single-diffractive dissociation process, $b+a \rightarrow b+X$, in
hadron-hadron collisions.}
\label{Fig:DR3R}
\end{SCfigure}

\begin{SCfigure}[][h]
\centering
\resizebox{0.2\textwidth}{!}{%
\includegraphics{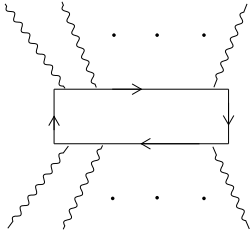}
}
\caption{Eikonal-type structure assumed for multi-reggeon interaction vertices.}
\label{Fig:Vrt}
\end{SCfigure}

\vspace{-0.3 cm}
\label{sec:1}
It was proposed in Ref. \cite{SDXS} to describe soft single-diffractive process in 
the model where any number pomeron exchanges is taken into account together with each 
reggeon (pomeron or secondary-reggeon, which was considered as $f$-trajectory) of the 
triple-reggeon diagrams. The screening corrections between initial and final hadrons 
were also taken into account in the eikonal approximation, as shown in Fig.~\ref{Fig:DR3R}. 
This approach allowed to describe inclusive spectra  in single diffractive $pp$ and $p\bar{p}$  
interactions from FNAL, ISR to Tevatron energies. 
For multi-reggeon interaction vertices eikonal-type structure was assumed, 
as shown in Fig.~\ref{Fig:Vrt}, which allowed to calculate and to sum the 
matrix elements. In this approach the system of hadrons, produced in diffraction 
dissociation of a hadron can be considered as a non-diffractive interaction of a 
hadron with the $q\bar{q}$ system which is responsible for inelastic interaction of 
reggeons or pomerons. 
The cross-section corresponding to the diagram shown in Fig.~\ref{Fig:DR3R}
was calculated using Gribov's reggeon calculus and has the following form:
\begin{eqnarray}
\frac{d \sigma}{d\zeta }=
\frac{1}{2}
\sum_{i,j,k = P,f} G_{ijk}
\int d{\bf b} d{\bf b_{1}}
\Gamma^{i}(\zeta_{2}, {\bf b_{2}})
\Gamma^{j}(\zeta_{2}, {\bf b_{2}}) \nonumber \\
\times
\Gamma^{k}(\zeta, {\bf b_{1}})
\exp\{-2\Omega_{1}(\xi, {\bf b})\}
\label{eq:sigma3R}
\end{eqnarray}
Where the following notations are introduced:
\begin{eqnarray}
\Gamma^{f}(\zeta, {\bf b}) = \frac{g_{0}^{f}}{\lambda_{0}^{f}}
\exp \left\lbrace (\alpha_{f}-1)\zeta - \frac{{\bf b}^{2}}{4\lambda_{0}^{f}}
-\Omega_{0}(\zeta, {\bf b}) \right\rbrace,  \label{eq:sigma3R1} \nonumber \\
\Gamma^{P}(\zeta, {\bf b}) = 1-e^{-\Omega_{0}(\zeta, {\bf b})},  \,  \,
\lambda_{i}^{k}=R_{i}^{k2}+\alpha_{k}^{\prime}\zeta, \nonumber\\
\Omega_{i}(\zeta, {\bf b}) = \frac{g_{i}^{P}}{\lambda_{i}^{P}}
\exp \left\lbrace \Delta \zeta - \frac{{\bf b}^{2}}{4 \lambda_{i}^{P}} \right\rbrace, \, \,
\xi=ln(s/s_{0}), \nonumber\\
\zeta=ln(M^{2}_{X}/s_{0}), \,  \, \zeta_{2}=\xi -\zeta,  \,  \, {\bf b_{2} = b -b_{1}}. \nonumber
\end{eqnarray}
The values of parameters are found from fit to elastic and inelastic diffractive data and 
have the following values: $\alpha_{f}$ = 0.7, $\alpha_{f}^{\prime}$ = 0.8 GeV$^{-2}$, 
$\Delta$ = 0.117,$\alpha_{P}^{\prime}$ = 0.25 GeV$^{-2}$, $g_0^P$ = 1.16 GeV$^{-2}$, 
$g_1^P$ =1.87 GeV$^{-2}$, $g_0^f$ = 8.2 GeV$^{-2}$, $g_1^f$=3.5 GeV$^{-2}$, $R_0^{P2}$ 
= 1.9 GeV$^{-2}$, $R_1^{P2}$ = 2.8 GeV$^{-2}$, $R_0^{f2}$ = 1 GeV$^{-2}$, $R_1^{f2}$ = 
1.8 GeV$^{-2}$, $G_{PPP}$ =0.0098 GeV$^{-2}$, $G_{PPR}$ =0.03 GeV$^{-2}$, $G_{RRP}$ =
0.005 GeV$^{-2}$, $G_{RRR}$ = 0.05 GeV$^{-2}$, $G_{PRP}$ = 0.013 GeV$^{-2}$, $G_{PRR}$ 
= 0.033 GeV$^{-2}$. $s_0$ is a constant scale factor, chosen to be $s_0$ = 1 GeV$^2$.

Spectra of particles produced in single-diffractive dissociation we calculate within the 
framework of QGSM. In the next section we give brief description of the model.

\vspace{-0.4 cm}
\section{QGSM}
\vspace{-0.3 cm}
In QGSM the production of a particle is defined through production of showers and each 
shower corresponds to the cut-pomeron pole contribution in the elastic scattering amplitude.  
The association of the pomeron with cylinder-type diagrams leads to the fact that in a single 
cut-pomeron diagram there are two chains of particles (strings). Analogously, in case of $n$ 
cut-pomerons there are 2$n$ chains. 
Inclusive distribution of particles in rapidities for non-diffractive events can be 
expressed in terms of rapidity distributions for 2$n$-chains $f_n(\xi,y)$:
\begin{equation}
\frac{d \sigma(\xi)}{dy }= \sum_{n} \sigma_{n}(\xi) f_{n}(\xi,y).
\label{Eq:RapDist}
\end{equation}
Here $\sigma_n(\xi)$ is the cross-section of producing 2$n$-chains and $\xi$~=~$\ln(s/s_0)$. 

Being interested in the spectra of particles produced in single diffractive process 
we calculate $\sigma_n$ for single-diffraction dissociation in the same way as for 
inelastic process in  $q\bar{q}-p$ interaction with $s=M^2$  or $\xi=\zeta$. For eikonal 
approximation, which we use the expression for $\sigma_n$ are well known \cite{QGSM}.  

Spectra of secondary particles in single-diffractive dissociation process integrated 
over $M^2$ are given by the following expression: 
\begin{equation}
\frac{d \sigma}{dy }= \sum_{n}
\int d\zeta \frac{d \sigma_{n}}{d\zeta }f_{n}(\zeta,y)
\label{Eq:RapDistSD}
\end{equation}
In order to calculate the distribution of hadrons produced during the fragmentation 
of the strings one should take into account that the hadron can be produced in each 
of the 2$n$-chains and the (di-)quarks, which stretch the strings, carry only a 
fraction of energy of the incoming protons. In these terms, the inclusive spectra can 
be written as convolution of the probabilities to find a string with certain rapidity 
length and the fragmentation function, which defines the distribution of hadrons in 
the string breaking process.

For meson-nucleon collisions the function $f_n(\xi,y)$ is written as follows \cite{QGSM}:
\begin{eqnarray}
f_n^h(\xi,y)=a^h\left[F^{h(n)}_{\bar{q}}(x_+)F^{h(n)}_q(x_-)
+F^{h(n)}_{qq}(x_+)F^{h(n)}_q(x_-) \right.\nonumber \\
\left.+2(n-1)F^{h(n)}_s(x_+) F^{h(n)}_{\bar{s}}(x_-)\right], 
\label{Eq:fn}
\end{eqnarray}
where $x_{\pm} = (1/2)[\sqrt{x_T^2+x^2} \pm x]$, $x_T=2m_T^h/\sqrt{s}$ and $x$ is the 
Feynman-$x$ of produced hadron $h$.
Here $\sqrt{s}$ is the center of mass energy of meson-nucleon system. $a^h$ is the 
density of hadrons $h$ produced at mid-rapidity in a single chain and its value is 
determined from experimental data.  In fact these are the only free parameters in 
the model that are fixed from fit to data. In articles \cite{QGSM} and \cite{QGSMKandP} 
the following values are found for them: $a^{\pi}$ = 0.44, $a^K$ = 0.055, $a^{p}$ = 0.07. 
The first two terms in Eq.~(\ref{Eq:fn}) correspond to the chains, which connect ``valence'' 
quarks and di-quarks, and the last term corresponds to chains connected to the sea 
quarks-antiquarks. The functions $F_i^{h(n)}(x)$ being a convolution of the structure 
$\psi(x)$, and fragmentation $G(x)$, functions have the following form for the quarks 
and diquark of the incoming proton:
\begin{eqnarray}
F_{q_{val}}^{h(n)}(x_{\pm})=
\frac{2}{3}\int_{x_{\pm}}^{1}dx\psi_{u_{val}}(n,x)G_{u}^{h}\left(\frac{x_{\pm}}{x}\right) \nonumber\\
+\frac{1}{3}\int_{x_{\pm}}^{1}dx\psi_{d_{val}}(n,x)G_{d}^{h}\left(\frac{x_{\pm}}{x}\right).\nonumber\\
F_{qq}^{h(n)}(x_{\pm}) =
\frac{2}{3}\int_{x_{\pm}}^{1}dx\psi_{ud}(n,x)G_{ud}^{h}\left(\frac{x_{\pm}}{x}\right) \nonumber \\
+\frac{1}{3}\int_{x_{\pm}}^{1}dx\psi_{uu}(n,x)G_{uu}^{h}\left(\frac{x_{\pm}}{x}\right).\nonumber\\
F_{q_{sea}}^{h(n)}(x_{\pm}) =\frac{1}{2+\delta}\left[
\int_{x_{\pm}}^{1}dx\psi_{u_{sea}}(n,x)G_{u}^{h}\left(\frac{x_{\pm}}{x}\right)\right.\nonumber\\
+\int_{x_{\pm}}^{1}dx\psi_{d_{sea}}(n,x)G_{d}^{h}\left(\frac{x_{\pm}}{x}\right)+\nonumber\\
\left.\delta \int_{x_{\pm}}^{1}dx\psi_{s_{sea}}(n,x)G_{s}^{h}\left(\frac{x_{\pm}}{x}\right)\right].\nonumber
\end{eqnarray}
Where $\delta$ is the strangeness suppression parameter ($\delta \approx 1/3$). Convolution functions 
for quarks and anti-quarks from incoming meson can be written analogously. 

In the model, the structure and fragmentation functions are determined by the corresponding 
Regge asymptotic behaviors in the regions $x \rightarrow 0$ and $x \rightarrow 1$ and for the 
full range of $x$ an interpolation is done. The structure functions of a proton are parameterized 
as follows \cite{QGSMKandP}:
\begin{eqnarray}
\label{Eq:StructP}
\psi_{d_v}(x,n) = \psi_{d_s}(x,n) = C_{d}x^{-\alpha_{R}}(1-x)^{\alpha_{R}-2\alpha_B+n}     \nonumber\\
\psi_{u_v}(x,n) = \psi_{d_s}(x,n) = C_{u}x^{-\alpha_{R}}(1-x)^{\alpha_{R}-2\alpha_B+n-1}   \nonumber\\
\psi_{ud}(x,n) = C_{ud}x^{\alpha_{R}-2\alpha_B}(1-x)^{\alpha_{R}+n-1}  \\
\psi_{uu}(x,n) = C_{uu}x^{\alpha_{R}-2\alpha_B+1}(1-x)^{\alpha_{R}+n-1}  \nonumber\\
\psi_{s}(x,n) = C_{s}x^{-\alpha_{R}}(1-x)^{2(\alpha_{R}-\alpha_B)-\alpha_{\phi}+n-1}\nonumber
\end{eqnarray}
Where $\alpha_{R}$=0.5, $\alpha_B$=-0.5. For meson structure function we will use the following 
parameterization 
\begin{eqnarray}
\psi_{d_v}(x,n) = \psi_{\bar{d}_v}(x,n) = \psi_{u_v}(x,n) = \psi_{\bar{u}_v}(x,n) = \nonumber\\
\psi_{s}(x,n) = C_{q}x^{\alpha_{R}}(1-x)^{\alpha_{R}+n-1}
\label{Eq:StructMeson}
\end{eqnarray}
In (\ref{Eq:StructP}) and (\ref{Eq:StructMeson}) $C_i$ are determined from normalization condition:
\begin{eqnarray}
\int_0^1 \psi_{i}(x,n) dx =1.\nonumber
\end{eqnarray}
General technique of constructing fragmentation functions is presented in Ref. \cite{QGSMfragm}. 
For instance, for fragmenting to a pion the parameterizations are (see \cite{QGSM} and \cite{QGSMKandP}):
\begin{eqnarray}
G_{d}^{\pi^{+}}(z) = G_{u}^{\pi^{-}}(z) = (1-z)^{2-3\alpha_{R}+\lambda}\nonumber\\
G_{u}^{\pi^{+}}(z) = G_{d}^{\pi^{-}}(z) = (1-z)^{ -\alpha_{R}+\lambda}\nonumber\\
G_{s}^{\pi^{+}}(z) = G_{s}^{\pi^{-}}(z) = (1-z)^{1 -\alpha_{R}+\lambda}\nonumber\\
G_{ud}^{\pi^{+}}(z)= G_{ud}^{\pi^{-}}(z) = (1-z)^{\alpha_{R}+\lambda-2\alpha_R}(1-z+z^{2}/2)\nonumber\\
G_{uu}^{\pi^{+}}(z) = (1-z)^{\alpha_{R}+\lambda-2\alpha_B}\nonumber\\
G_{uu}^{\pi^{-}}(z) = (1-z)^{\alpha_{R}+\lambda-2\alpha_B+1}\nonumber
\end{eqnarray}
We do not list fragmentation functions for kaons and (anti-)protons, they are taken from Ref. \cite{QGSMKandP}.

\section{Numerical results}
\begin{figure}[h]
\begin{center}
\resizebox{0.45\textwidth}{!}{%
\includegraphics{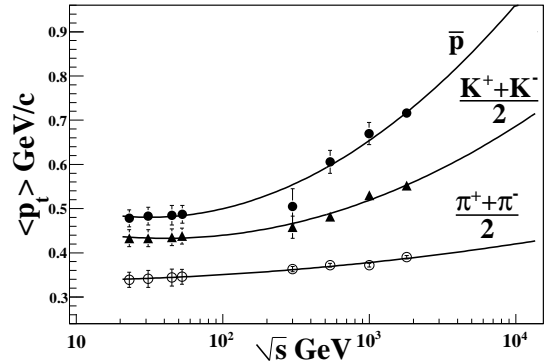}}
\caption{
The average transverse momentum $<p_t>$ as a function of $\sqrt{s}$ 
for $\pi^{\pm}$, $K^{\pm}$, $\bar{p}$ \cite{pt}. The $\pi$ and $K$ data are averaged 
over positive and negative particles.}
\label{Fig1:ptVsCMSE}
\end{center}
\end{figure}

\vspace{-0.3 cm}
There are not much data on spectra of secondary particles produced in soft diffraction 
processes. Only available data are on charged particles pseudorapidity distribution in 
single-diffractive dissociation process. QGSM has been used successfully to describe data 
on secondary hadron inclusive cross-section integrated over transverse momentum, such as 
rapidity and multiplicity distribution, ratio of particles and it explains many characteristic 
features of hadron-hadron soft interactions. But for calculating pseudorapidity distribution 
from Eq. (\ref{Eq:RapDist}) or (\ref{Eq:RapDistSD}) we must know mean transverse momentum of 
particles. For this purpose we used experimentally measured $<p_t>$ for different types of 
particles  (here we assume that the sample of secondary charged particles consists from 
$\pi^{+}$, $K^{+}$ and protons, and their anti-particles). A successful fit to the data from 
ISR to Tevatron energies is achieved with the second-order polynomial function of $\ln(s)$. 
The result of the fit is presented below and compared with data in Fig.~\ref{Fig1:ptVsCMSE}.
\begin{eqnarray}
\label{Eq:ptVsCMSE}
<p_t^{\pi}> = 0.34 -0.002 \ln s + 0.00035 \ln ^{2} s,\nonumber\\
<p_t^{K}> = 0.55 -0.031 \ln s + 0.001  \ln^{2} s,\\
<p_t^{p}> = 0.65 -0.045 \ln s + 0.0036   \ln^{2} s.\nonumber
\end{eqnarray}
In the following for each mass of the diffractive system we calculate $<p_t>$ of each 
particle type using (\ref{Eq:ptVsCMSE}).

\begin{figure}[h]
\begin{center}
\resizebox{0.5\textwidth}{!}{%
\includegraphics{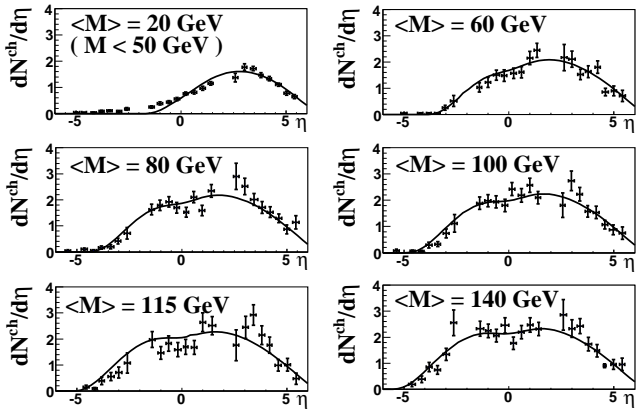}
}
\caption{
Description of UA4 data on charged particles pseudorapidty distribution in $p\bar{p}$ 
single-diffractive evens for different values of the diffractive mass.}
\label{Fig:UA4}
\end{center}
\end{figure}

UA4 \cite{UA4} provides data on charged particles pseudo-rapidity distribution in 
single-diffractive $p\bar{p}$ interaction at $\sqrt{s}$ = 546 GeV for six intervals 
of the mass for diffractively produced system. In Fig.~\ref{Fig:UA4} we compare the 
results of theoretical calculation with these data. For each mass bin the theoretical 
curve is calculated at mean value of the system’s mass, as indicated in each graph.
 
\begin{figure}[h]
\begin{center}
\resizebox{0.41\textwidth}{!}{%
\includegraphics{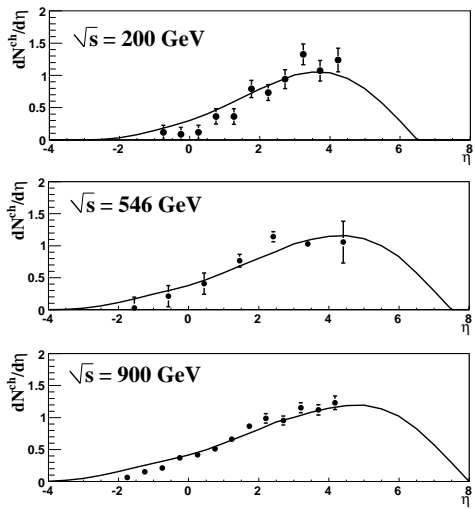}
}
\caption{
Description of UA5 data on charged particles pseudorapidity distribution in $p\bar{p}$
single-diffractive evetns. The indicated errors are statistical and the systematical 
errors are unknown.}
\label{Fig:UA5}
\end{center}
\end{figure}

UA5 \cite{UA5} measured charged particles pseudorapidity distribution in $p\bar{p}$ 
single-diffractive interaction at three different energies ($\sqrt{s}$ = 200 GeV, 546 
GeV and 900 GeV), and for integrated mass spectra of the diffractive system (
$M^2 \leq 0.05s$
). 
In Fig.~\ref{Fig:UA5} we give description of these data. 
\section{Conclusion}
Incorporating the model, proposed in Ref~\cite{SDXS} for description of cross sections 
for single-diffractive dissociation process, with the Model of Quark-Gluon Strings, a 
good description of available S$p\bar{p}$S data on particles spectra in $p\bar{p}$  
single-diffractive dissociation process was obtained. We stress that we had no free 
parameters in this analysis. We hope that this approach will give a reliable predictions 
for both cross sections of diffraction dissociation to large mass states and particle 
production in this process at LHC energies and can be used to calculate backgrounds from 
diffractive processes in experiments where direct observation of diffractive processes 
will not be possible.

\section{Acknowledgements}
The work of A.B.K. was partially supported by the grants RFBR 0602-72041-MNTI, 0602-17012, 
0802-00677a and Nsh-4961.2008.2.

\end{document}